\documentclass{article}
\title{Correspondence Between Classical and Quantum Theory by $f$-Deformed Coherent State}
\author{R. Roknizadeh$^{1,3}$, S. A. A. Ghorashi$^2$ and H. Heydari$^3$\\
\small{ 1. Department of Physics, Quantum Optics Group, University of Isfahan, Isfahan, Iran}\\
\small{2. Department of Physics, University of Houston, Houston, TX 77204-5506, USA}\\
\small{3. Physics Department, Stockholm University 10691 Stockholm Sweden}}
\begin{document}
\maketitle
\begin{abstract}
 Generalized $f$-coherent state approach in deformation quantization
framework is investigated by using a $\ast $-eigenvalue equation. For this
purpose we introduce a new Moyal star product called $f$-star product, so
that by using this ${\ast} _{f}$-eigenvalue equation one can obtain exactly
the spectrum of a general Hamiltonian of a deformed system.
\end{abstract}
\section{Introduction}
 A deformation of a mathematical object (here a path in phase space)
is a family of the same kind of objects depending on some parameter(s). The
deformation of algebra of functions is central to the deformation quantization method.
 Deformation quantization is introduced in \cite{1} and extensively
developed during recent years (for a survey see \cite{2}).

   Since the appearance, the deformation quantization method is attracting an ever-growing attention and more than ever it has found  many application in different fields of physics and mathematical physics \cite{3,4} such as high energy physics, cosmology \cite{5}, string theory \cite{6} and brane theory \cite{7}.

  In other side, different generalization of coherent states are used
extensively in different methods of quantization, such as geometric
quantization \cite{8} and specially Berezin quantization \cite{9}, as
quasi-classical states to relate the classical phase space to quantum
mechanical Hilbert space. Because of their nonclassical properties such as
amplitude squeezing, quantum interference and sub-Poissonian statistics
\cite{10}, these states attract also many attentions in different fields of
physics specially quantum optics \cite{11}.

  Our purpose in this paper is introducing a new deformation
quantization scheme using a special kind of generalized coherent states,
which in quantum optics literature known as nonlinear or $f$-coherent states
 \cite{12}. In this article we exploit the introduced $\ast $-eigenvalue
equation by Fairlie et al. \cite{13}, for obtaining  the energy spectrum of a general Hamiltonian.\\

\section{The $f$-deformed coherent states}

Man'ko et al., \cite{12}, have introduced (nonlinear) coherent states of an $f$%
-deformed algebra as right hand eigenstates of the generalized annihilation operator $\hat{A}=\hat{a%
}f(\hat{n})$, where $\hat{n}=\hat{a}^{\dag }\hat{a}$ is the usual number
operator. These $f$-coherent states can be expressed as
\begin{equation}  \label{20}
\hat{A}\left\vert \zeta ,f\right\rangle =\zeta \left\vert \zeta
,f\right\rangle, \quad \hat{A}^{\dag }=f\left( \hat{n}\right) \hat{a}^{\dag }.
\end{equation}
Thus, corresponding commutation relations are calculated easily as
\begin{equation}  \label{22}
\left[ \hat{A},\hat{A}^{\dag }\right] =\left( \hat{n}+1\right) f^{2}\left(
\hat{n}+1\right) -\hat{n}f^{2}\left( \hat{n}\right) ,\quad \left[ \hat{n},\hat{A}\right] =-\hat{A},\quad\left[ \hat{n},\hat{A}^{\dag }\right]
=\hat{A}^{\dag }.
\end{equation}

Normalized NCSs can be expanded in the terms of Fock states as
\begin{equation}  \label{24}
\left\vert \zeta ,f\right\rangle =N_{f}\sum_{n=0}^{\infty }\frac{ \zeta ^{n}%
}{\sqrt{n!}f(\hat{n})!}\left\vert n\right\rangle .
\end{equation}
\section{$f$-star product}
The Moyal star product were used in several different systems and,  as is usual in different domain of physics, 
harmonic oscillators are most important example. After defining  two following functions
$a=\frac{q+ip}{\sqrt{2}}, \bar{a}=\frac{q-ip}{\sqrt{2}}$, 
the classical Hamiltonian of SHO can be written as
$H=\omega \bar{a}a.$
We can rewrite the star product in terms of the functions $a,\bar{a}$
as
\begin{equation}  \label{11}
\ast _{M}=\exp \left[ \frac{\hbar }{2}\left( \overleftarrow{\partial _a}
\overrightarrow{\partial_{\bar{a}}}-\overleftarrow{\partial_{\bar{a}}}
\overrightarrow{\partial _a}\right) \right] .
\end{equation}
Therefore, $\ast _{M}-$commutator algebra of harmonic oscillator is given by
\begin{equation}  \label{12}
\frac{1}{\hbar }\left[ a,\bar{a}\right] _{\ast _{M}}=\frac{1}{\hbar }(a\ast
_{M}\bar{a}-\bar{a}\ast _{M}a)=1.
\end{equation}
In the past years many attempts were devoted to
generalization of this product for systems containing damping \cite{17}, and
coupling to other system and anharmonicity are incorporated in Hamiltonian
 \cite{18}. Here we propose exploring the generalized coherent states to
extend deformation quantization for these systems via introducing a new
generalization of the Moyal star product. An explicit differential form of
star product for these systems is yet not introduced, although some attempts
are done \cite{19}.

  In this work we will show that our new star product, we call
 $f$-star product, can preserve the $\ast _{f}-$commutator of the deformed
annihilation and creation functions, $A=a f(n)$ and $\bar{A}=f(n)\bar{a%
}$, similar to commutation relation of annihilation and creation function
that preserved Moyal-$\ast _{M}-$commutator.
We introduce now the $f$-star product as
\begin{eqnarray}
\ast _{f} &=&\exp \left[ \frac{\hbar }{2}\left( \left( n+1\right)
f^{2}\left( n+1\right) -nf^{2}\left( n\right) \right) \left( \overleftarrow{%
\frac{\partial }{\partial A}}\overrightarrow{\frac{\partial }{\partial \bar{A%
}}}-\overleftarrow{\frac{\partial }{\partial \bar{A}}}\overrightarrow{\frac{%
\partial }{\partial A}}\right) \right]  \label{25} \\
&=&\exp \left[ \frac{i\hbar }{2}\left( F(n)\right) \left( \overleftarrow{%
\frac{\partial }{\partial q}}\overrightarrow{\frac{\partial }{\partial p}}-%
\overleftarrow{\frac{\partial }{\partial p}}\overrightarrow{\frac{\partial }{%
\partial q}}\right) \right],  \nonumber
\end{eqnarray}
where $F(n)=\left( \left( n+1\right) f^{2}\left( n+1\right)
-nf^{2}\left( n\right)\right)/f(n)f(n+1)$. This $f$-star product has all
properties of a star product and according to \cite{20} can be written as
\begin{eqnarray}\label{26}
k\ast_f g &=&kg+\frac{i\hbar }{2}F(n)\left[ \frac{\partial k}{\partial q}\frac{%
\partial g}{\partial p}-\frac{\partial k}{\partial p}\frac{\partial g}{%
\partial q}\right] \\
&&-\frac{\hbar^2 }{4}k\left[ F(n)\left( \overleftarrow{\frac{\partial }{%
\partial q}}\overrightarrow{\frac{\partial }{\partial p}}-\overleftarrow{%
\frac{\partial }{\partial p}}\overrightarrow{\frac{\partial }{\partial q}}%
\right) F(n)\left( \overleftarrow{\frac{\partial }{\partial q}}%
\overrightarrow{\frac{\partial }{\partial p}}-\overleftarrow{\frac{\partial
}{\partial p}}\overrightarrow{\frac{\partial }{\partial q}}\right) \right]
g+O(\hbar ^{3}), \nonumber
\end{eqnarray}
for arbitrary functions $k$ and $g$ on phase space.

Based on the Kontsevich work \cite{21}, this star product with $F(n)$, which
depends on $q$ and $p$, keeps associativity  up to the second order, ${\cal{O}}(\hbar ^{2})$.
One can easily see that differential operators act on $F(n)$ just in the 
terms of  second order in  $\hbar $. In the following sections, we will preserve the first order of $%
\hbar $ to obtain the $\ast_{f}$-eigenvalue equation. Thus, we
can simply extract $F(n)$ from product expansion and one can just apply
differential operators on functions $k$ and $g$.

Now, we can easily show the following correspondence between two algebras
\begin{eqnarray}
\left[ \hat{A},\hat{A}^{\dag }\right] &=&\left( \hat{n}+1\right) f^{2}\left(
\hat{n}+1\right) -\hat{n}f^{2}\left( \hat{n}\right) ,  \label{27}\\
\frac{1}{\hbar }\left[ A,\bar{A}\right] _{\ast _{f}} &=&\left[ \left(
n+1\right) f^{2}\left( n+1\right) -nf^{2}\left( n\right) \right] .  \nonumber
\end{eqnarray}
The first equation corresponds to commutation relation between generalized
creation and annihilation operators whereas the second one is related to the  generalized
functions. It is obvious in the limiting case $f(n)$ $\rightarrow 1$, $f$-star
product will be identified with Moyal star product and
$\ast _{f}$-commutator identified with $\ast _{M}$-commutator.

  Now, by calculating the Wigner function of generalized coherent
state and using the new defined star product we can obtain the corresponding
$\ast $-eigenvalue problem. The  diagonal elements of Wigner function in
this representation, according to \cite{12}, are given by,
\begin{equation}
W^{f}(q,p) =\sum_{n=0}^{\infty }W_{nn}^{f}(q,p)=2N_{f}^{2}e^{-\left(
q^{2}+p^{2}\right) }\sum_{n=0}^{\infty }\frac{\left( -1\right) ^{n}}{n!\left[
f(n)!\right] ^{2}}\left\vert \zeta \right\vert ^{2n}L_{n}(2\left(
q^{2}+p^{2}\right) ). 
\end{equation}
 Using the following phase space Hamiltonian as ansatz, we obtain the corresponding $\ast$-eigenvalue equation,
 \begin{eqnarray}
H &=&\frac{\hbar \omega }{2}\left( \bar{A}A+A\bar{A}\right)  \label{29}\\
E_n &=&\frac{\hbar \omega }{2}\left( \left( n+1\right) f^{2}\left( n+1\right)
+nf^{2}\left( n\right) \right)  \label{29-1} \\
&=&\frac{\hbar \omega }{2}\left[ \left( \frac{\left( q^{2}+p^{2}\right) }{2}%
+1\right) f^{2}\left( \frac{\left( q^{2}+p^{2}\right) }{2}+1\right) +\left(
\frac{\left( q^{2}+p^{2}\right) }{2}\right) f^{2}\left( \frac{\left(
q^{2}+p^{2}\right) }{2}\right) \right].\nonumber
\end{eqnarray}
By putting this Hamiltonian in $\ast _{f}-$eigenvalue equation, the desired equation will be achieved,  
\begin{eqnarray}
H\ast _{f}W_{n}^{f} &=&E_{n}W_{n}^{f}  \label{30} \\
&&+\frac{i\hbar }{2}\left( \frac{\left( n+1\right) f^{2}\left( n+1\right)
-nf^{2}\left( n\right) }{f(n)f(n+1)}\right) \left( \frac{\partial H}{\partial q}\frac{\partial W}{\partial p}-\frac{\partial H}{\partial p}\frac{\partial W}{\partial q}\right).  \nonumber 
\end{eqnarray}
The first term in the right hand side is exactly the spectrum of the Hamiltonian
of a $f-$deformed harmonic oscillator. By some simple consideration, we can
show that the imaginary part is equal to zero.\\

For more detail and  some applications of this method to different examples, see \cite{22}.

{\bf Acknowledgments}\\

R. R. wishes to thank Vice President of Research and Technology  of University of Isfahan, for financial support  and Albanova University Center of Stockholm for hospitality during sabbatical leave.  H. H. thanks  the Swedish Research Council (VR) for financial support. 


\begin{thebibliography}{0}
\bibitem{1} F. Bayen, M. Flato, C. Fronsdal, A. Lichnerowicz and D. Sternheimer,
\emph{Ann. Phys., NY} \textbf{111}, 61 (1978).

\bibitem{2} M. Bordemann, \emph{Journal of Physics: Conference Series} \textbf{103}, 012002 (2008).

\bibitem{3} H. Bursztyn, S. Waldmann, \emph{Lett. Math. Phys.}
\textbf{72}, 143 (2005).

\bibitem{4} A. C. Hirshfeld, P. Henselder, \emph{Annals of Physics} \textbf{302}, 5977 (2002).

\bibitem{5} R. Cordero, H. Garcia-Compean, F. J. Turrubiates, \emph{Phys. Rev. D} 
\textbf{83}, 125030 (2011).

\bibitem{6} H. Garcia-Compean, J.F. Plebanski, M. Przanowski, F.J. Turrubiates, \emph{J.
Phys. A: Math. Gen.} \textbf{33}, 7935 (2000).

\bibitem{7} L. Cornalba, R. Schiappa, \emph{Commun. Math. Phys.} \textbf{225}, 33 (2002).

\bibitem{8} N. M. J. Woodhouse, \textit{Geometric Quantization}, (Clarendon
Press Oxford, 1992).

\bibitem{9} F. A. Berezin, Quantization, \emph{Math. USSR-Izv.} \textbf{8}, 1109 (1974).

\bibitem{10} C. K. Hong and L. Mandel, \emph{Phys. Rev. Lett.}  \textbf{54}, 323 (1985),
M. C. Teich and B. E. A. Saleh, Quantum Opt.  \textbf{1}, 153 (1989).

\bibitem{11} J. R. Klauder and B.S. Skagerstam, \textit{Coherent States,
Applications in Physics and Mathematical Physics} ( World Scientific Singapore 1985).

\bibitem{12} V. I. Man'ko, G. Marmo, E. C. G. Sudarshan and F. Zaccaria, \emph{Phys. Scr.}
\textbf{55}, 528 (1997).

\bibitem{13} D. Fairlie and C. Manogue, \emph{J. Phys. A: Math. Gen.} \textbf{24} 3807 (1991), T. Curtright, D. Fairlie and C. Zachos, \emph{Phys Rev D} \textbf{58}, 025002 (1998).

\bibitem{17} G. Dito, F.J. Turrubiates, \emph{Phys Lett A} \textbf{352}, 309 (2006).

\bibitem{18} F. Becher, N. Neumaier, S. Waldmann, \emph{Lett. Maht. Phys.} ‎\textbf{92}‎, 155 (2009).

\bibitem{19} V. I. Man'ko, G. Marmo, E. C. G. Sudarshan, F. Zaccaria, \emph{Phys. Lett. A}  \textbf{372}, 4364 (2008).

\bibitem{20} A. Stern, \emph{Nucl. Phys. B} \textbf{754}, 236 (2006).

\bibitem{21} M. Kontsvich, \emph{Lett. Math. Phys.} \textbf{66}, 157 (2003).

\bibitem{22}  S. A. A. Ghorashi, R. Roknizadeh, M. Bagheri Harouni, \emph{Int. J.  Mod.  Phys.}, \textbf{27},   1250095 (2012).
\end{thebibliography}
\end{document}